# The Case For Alternative Web Archival Formats To Expedite The Data-To-Insight Cycle

Xinyue Wang
xw0078@vt.edu
Virginia Polytechnic Institute and State University
Blacksburg, VA

Zhiwu Xie
zhiwuxie@vt.edu
Virginia Polytechnic Institute and State University
Blacksburg, VA

## ABSTRACT
The WARC file format is widely used by web archives to preserve collected web content for future use. With the rapid growth of web archives and the increasing interest to reuse these archives as big data sources for statistical and analytical research, the speed to turn these data into insights becomes critical. In this paper we show that the WARC format carries significant performance penalties for batch processing workload. We trace the root cause of these penalties to its data structure, encoding, and addressing method. We then run controlled experiments to illustrate how severe these problems can be. Indeed, performance gain of one to two orders of magnitude can be achieved simply by reformatting WARC files into Parquet or Avro formats. While these results do not necessarily constitute an endorsement for Avro or Parquet, the time has come for the web archiving community to consider replacing WARC with more efficient web archival formats.

## 1 INTRODUCTION
The Web has evolved to become a primary carrier of our knowledge and memory. It is, however, inherently decentralized, dynamic, and ephemeral. Despite valiant efforts [17, 54] to inject a sense of temporality in the web architecture [29], the majority of today's web still focuses on disseminating the most up-to-date information, and does not leverage any built-in mechanism to persist representations of web resources that are either purposefully retired, replaced, or vanishing due to negligence. As a consequence, increasingly more hyperlinks break or no longer point to their intended representations [31, 50]. This phenomenon, referred to as "link rot", endangers the lineage of our history.

To battle this loss, web archives are created to collect and preserve the current web content for potential future reuse. A growing number of cultural heritage institutions actively engage in web archiving. A 2011 survey reported 42 web archiving initiatives around the world [23]. The number has increased to 68 in 2015 [15]. A more recent survey reported 119 web archiving programs in the United States alone [18], 61% of which were part of university libraries or archives. Large-scale, comprehensive web archiving initiatives include the Internet Archive [8], the Common Crawl [16], and many programs at national libraries and archives. These initiatives have preserved large amounts of web content, some dating as far back as in the 1990s. As of May 2019, the Internet Archive alone has archived about 40 petabytes of web content, including 362 billion web pages and 754 billion URLs, and is growing at the speed of more than 5000 URLs per second [1].

Similar to search engines, the vast majority of the content in web archives are automatically collected by web crawlers. Unlike search engines, however, the collected content is not primarily intended for immediate consumption, therefore with no rush to be fully indexed and made available for search and browsing through a highly available and highly scalable platform like Google. Instead, the HTTP requests from the crawlers and the responses sent back by the origin servers are concatenated and aggregated into larger files intended to be stowed away for safekeeping in archival storage. The Web ARChive, or WARC [28] file format became an international standard in 2009. Archiving the web has since became synonymous with creating WARC files.

Some web archives provide limited browsing interfaces to WARC files. For example, the Internet Archive's Wayback Machine [9] allows users to retrieve and play back a historic web page using its original URL and a timestamp. Similar browsing tools have been developed and adopted elsewhere [2, 13, 56]. Browsing functions like these typically leverage CDX files [12], which point to the archived record and also contains metadata such as the URL, timestamp, MIME type, and response code, etc. to assist page-by-page browsing. Nevertheless, the central theme of today's web archives is still predominantly on web content collections and preservation. Reuse beyond the prescribed browsing pattern is rarely supported. As a result, the archival system architecture is centered around WARC files and not optimized for research-driven analytics workflows.

It is difficult to develop generalizations over web archives. Despite the significant growth of web archives in size, the whole web and its complete history grow even faster. The web archives therefore remain a rather thin sample. Most general purpose web archives tend to be sparse as well as shallow, capturing only a very small percentage of published versions [3, 4]. The sparsity grows even thinner when the web resource is addressed farther away from the home page. This is primarily due to the crawler-based collection method used predominantly by these archives and the associated resource, policy, and legal constraints. Given the low sampling rate of the archived web content, we need to exercise caution in generalizing and extrapolating the insights gained from close reading. In contrast, treating web archives as big data sources to extract statistics and explorative insights may be more profitable.

Reusing web archives for analytical research, however, usually require archive users to deploy big data systems such as Apache

---
[1]Private Communications

Hadoop or Apache Spark. These systems tend to be sensitive to performance, which can easily fluctuate by orders of magnitude. Performance may therefore determine whether it is even feasible to explore certain research questions. Although a plethora of factors may impact the performance of big data systems, domain experts and big data system designers do not always align their strategies. When approaching a domain problem, big data system designers typically refrain from questioning the domain common practices. Instead, these practices are taken as matter-of-fact inputs for design. However, the domain status quo only reflects best practices extracted from the past experience and requirements that may be irrelevant in big data settings. Although keenly aware of this, domain experts do not always heed the contexts and nuances behind the drastic performance gain demonstrated by many big data systems. If no adaptation is made when using these systems, the poor initial and boundary conditions introduced by unfavorable domain common practices may penalize performance. It is in this gap that we see opportunities, particularly concerning data formats. Indeed, many de facto domain-specific data formats are being scrutinized for their performance penalty when used in big data settings. This paper focuses on the web archival format. More specifically, we present the case for the web archiving community to move away from using WARC as the default archival format. We analyze the root causes on why WARC penalizes big data analytics performance in comparison to more efficient data formats, and run controlled experiments to quantify these losses. The results also provide insights to generalize the types of archival data format we should avoid for web archives.

## 2 REUSE WEB ARCHIVES AS BIG DATA SOURCES

In order to evaluate performance, we must first define typical workloads against which benchmarking will be performed. Reusing web archives as big data sources has been steadily gaining momentum in research areas such as history [10, 24, 42] and communications [55]. Lin et al. observes that the frontier of contemporary historians' research focuses usually moves along about 20 to 30 years before now [34]. They therefore anticipate that many historians will soon start to look at the history of the 1990s and suddenly enter an age of abundance [42]: web archives have preserved unprecedentedly large amounts of historic information.

How will future web archive users, historians included, explore web archives in addition to close reading? We can only speculate, although extrapolating from problems researchers are currently exploring should give us higher confidence in our educated guesses. We identify 3 types of representative workloads on web archives, described in the following.

Take a recent research study [36] as an example. Mallapragada focuses her research on interpreting Indian immigrants' homepages from the mid-1990s to mid-2000s, and the researcher has already developed a list of URLs considered representative. We may then speculate that she would be interested in knowing how many relevant snapshots have been taken by a web archive during her time period of interest. This is a simple query that can be easily answered by filtering only existing metadata (list of URLs and timestamp range) about the archived payload without needing to access the actual payload, referred hereafter as Type 1 workload.

Mallapragada also noticed that many homepages contain links named "about me" or advertising for international calling cards. Wouldn't it be useful to be able to tell the percentage of relevant homepages that have these features? In order to answer these questions using the Type 1 query, however, we will need to query against a "link" metadata field that provides information about all links going out from a webpage. This field does not already exist, therefore needs to be extracted from the HTML payload. Type 2 workload refers to the type of queries that only filter against existing metadata (in this case the same URLs and timestamps as above), but the ultimate goal is to be able to extract information from relevant payloads (in our case, extract links from the relevant Indian immigrants' homepages). No matter what kind of data processing is to be performed, the relevant records including their payloads must be loaded into memory first. For the sake of evaluating performance on equal ground, we throw away the information extraction portion of the Type 2 workload, and only consider the performance up till the relevant records have been loaded into memory. Type 2 workload is also useful for partitioning large archival data sets into manageable chunks. It is not quite realistic for a small research team with a shoestring budget to build a data processing system big enough to query 40 petabytes of archived data from the Internet Archive all at once. Using Type 2 queries, we may filter the most relevant resources out of much larger datasets, then output them as a new but smaller data set. In this particular case, we may output a new file from the memory containing only the relevant records, e.g., all Indian immigrants' homepages captured between the mid-1990s and the mid-2000s.

The smaller data set resulting from the above partition may then be fed into the Type 3 workload, where no further record filtering is necessary, and the main focus is to manipulate if not all then a high percentage of the records in the data set. Typically all records are loaded into memory. Typical data processing includes parsing and field extraction, text mining, and topic modeling, etc. [24]. In our case, the useful data manipulation involves link extractions.

There exists an important difference between Type 3 and Type 2 workload inclusive of data processing. Type 2 workloads must load into memory data extracted from much larger archival files stored on the disk. This usually involves lots of iterative random disk access, which is much slower than a sequential disk read typically used in a Type 3 workload. In light of this, if the data extracted from a Type 2 workload is expected to be used repeatedly for various purposes, it is usually cheaper to first persist the extracted data on disk and then move the smaller data set elsewhere for further processing. In other words, even if we can afford to build a query system that can hold all 40 petabytes of data, it is still more efficient to use this system only to partition the larger data set instead of directly handling batch analytics. Therefore, the suitable cyberinfrastructure pattern to reuse web archives should always be the "Bridge" [57], where archival files are stored separately from where the data are to be processed.

Under the "Bridge" pattern, however, we must assume we initiate a web archival analytics task from a "cold" system. That is, relevant data have not already been loaded into memory from disk for querying. As indicated above, the actual data loading can be a major performance bottleneck. For example, after all relevant data have been loaded into system memory, even if system A takes 1/10 time to perform the same analytics task as system B, we still cannot be sure system A outperforms B, because we do not know how much time each system needs to load the data into memory. It is therefore critical to evaluate the full "Data-To-Insight" performance, which starts from the moment full archival data have been successfully copied to the disk of a "cold" (e.g., with empty cache) big data system, and ends at the moment useful information has been or is ready to be extracted from the memory.

To summarize, this section outlines three types of workload we believe will be representative of how researchers will be reusing web archives as big data sources. These workload types are in agreement with the filter, analyze, aggregate, and visualize, or FAAV web archive workflow previously identified [34], although our focus is more on characterising the underlying data manipulation patterns than their purposes and utilities.

## 3 RELATED WORK

In this section we present related work that helps to justify our emphasis on archival data format.

The state of the art of processing big data exploit various patterns of the computing systems and algorithms for accessing and querying the data. For example, it is generally assumed that as the data size increases, its complexity remains largely unchanged. In other words, large datasets are comprised of structurally repetitive records with low interdependency. Large datasets can therefore be partitioned and replicated, and processed on shared-nothing commodity machines in parallel. Web archives fit perfectly into this pattern.

Although Apache Hadoop is one of the more successful implementations of this pattern, MapReduce style processing has long been used by parallel databases. Pavlo et al. [47] shows that although it takes much longer for a parallel DBMS to load raw data, once the data are in place, its query speed is much faster than Hadoop. This finding brings to the front the tradeoff between data loading and querying/processing. Here, data loading refers to the process of changing the layout of the original data to another one. The new layout is specified by the parallel DBMS and has been optimized for its querying performance. It naturally follows that if the original data is already laid out in the way specified by the parallel DBMS, then spending resources to reformat data would be unnecessary. Unfortunately, parallel DBMS systems, many of which were proprietary, had not thought of standardizing their data layouts as open file formats. The situation has since changed with the growth of Apache Hadoop and Apache Spark. Two standard data formats, Parquet [6] and Avro [7], have been developed from these projects and adopted as their default data persistence formats. Although not intended for this comparison, circumstantial evidence in [46] indicates that parsing data formatted in Parquet and Avro can be at least one order of magnitude faster than parsing the same data formatted in JSON.

Some domain experts have been vocal about changing domain-specific file formats for big data processing. For example, the astronomy community have had long discussions on limitations of the FITS format [25, 43, 53]. Although performance is on the agenda, the discussions tend to be surrounding qualitative features, e.g., maintaining human-readability [53]. Biologists have also complained about using HDF5 to store large microscopy data [5].

However, many researchers, particularly computer scientists working on domain problems, do not bother to advocate for data format changes. They take the status quo as is and attempt to improve performance from there. Alagiannis et al. [1] argue that with the rapid growth of research data, the performance bottleneck has moved from querying to data loading. It is therefore more advantageous to trade the query speed for the capability to perform queries directly on raw data *in situ*. If designed properly, the sub-optimal querying performance can be compensated by the elimination of data loading. This stance persisted in Karpathiotakis et al. [30], who adapted their querying system to the ROOT file format specified by CERN's Large Hadron Collider project. Similarly, gap-filing software have been developed to improve the I/O performance for using HDF5, netCDF [35, 58] or FITS [48] in a Spark environment. Further, many have worked towards improving the parsing performance of the inefficient yet widely adopted JSON format [32, 46]. These improvements leverage various hardware/software performance gaps [59] but can not fully eliminate the inherent inefficiency incurred by a low-performing data format. They can, however, be used to speed up the conversion from a low-performing data format to a more efficient one. A rare exception to this stoic stance comes from the ADAM project [37, 45]. Here, the SAM/BAM file formats used to store large genome datasets have been replaced by a combination of Avro and Parquet formats, which helped to achieve a 50x speedup using Apache Spark.

For web archiving applications, Warcbase [33] chooses to load WARC files into HBase for querying. As such, archived data will need to persistently occupy the analytical systems, as in the "Network" pattern [57]. This is not suitable for long-term preservation. ArchivesUnleashedToolkit [52], the successor of Warcbase, also took WARC file format for granted but did not assume the existence of the accompanying CDX. Querying an ArchivesUnleashedToolkit application therefore requires repeated loading of the WARC files in full. ArchiveSpark [27], on the other hand, leverages CDX to selectively load WARC. Without sophisticated I/O scheduling, however, a full disk scan can still outperform many selective disk reads bundled together. Both ArchivesUnleashedToolkit and ArchiveSpark perform data processing using Apache Spark. Their main difference is how they load files from the disk into memory. In our comparison, we will perform data processing using Spark SQL on top of Apache Spark. After data is loaded into memory, we assume the performance difference between Spark and Spark SQL should not be significant. Significant performance gain should therefore be attributed to the file format differences.

## 4 WARC'S PERFORMANCE DEFICIENCIES

WARC is essentially a text-based file format that linearly concatenates many web records. Each record is made up of a header and a content block. The header consists of multiple key-value pairs describing the record; the content block contains the payload data collected from the web; both are encoded in text. Even non-textual payloads such as images, audio/vidio, and application media types are transcoded into flat US-ASCII text [20]. The WARC specification borrows its record-level structural and formatting conventions heavily from existing text-based web messaging protocols, especially RFC 2616 [19] and RFC 2822 [49]. Much of the messages exchanged during HTTP transactions may be directly inserted into WARC records without modification. It is therefore easier and cheaper to create WARC files while crawling the web than parsing the HTTP responses and then writing into another format. However, the WARC format is not optimized for analytical work loads. In this section we characterize the root causes of its potential performance penalty. Quantitative analyses will be presented in the next section.

## 4.1 Data Structure: Linear Concatenation

Archival files are bulky and most likely have to be persisted on cheaper, slower "cold" storage media. They need to be brought into faster media, e.g., RAM, to be useful. Archival formats determine how archived information is laid out on persistent storage. A naive data structure built using the linear concatenation method specified by WARC, will penalize the reuse performance, because it tends to be slower to traverse, retrieve, and load useful information from such a data structure into memory.

WARC stacks a large number of records consecutively without indexing. Reusing WARC files typically involves reading the whole file and parsing the records consecutively into programmable data structures such as RDDs or Java/Python objects in RAM. WARC alone does not provide a mechanism to jump directly to a useful record without parsing all those before and after it. This changed when the CDX specification was introduced. A CDX file itself is another text file that linearly concatenates useful metadata, one line per record, in the hope that the metadata will help users to decide which record(s) will be useful. It also maintains the offset information for each record, which allows us to read individual records. With the CDX, many structured queries may be turned into the following sequence: 1) parse the CDX file, 2) query the parsed information to decide which records are useful, 3) using the offsets to retrieve the useful records individually, and 4) load each record into memory and parse it, including parsing again the metadata already parsed in step 1. Even with CDX, there are still abundant opportunities to improve the analytical performance. Besides metadata duplication, if many useful records are closely aggregated in nearby blocks of storage, it is sometimes faster to first read these blocks in full into the memory and then perform the queries, rather than following the above 4 steps sequentially. However, caching, cost estimates, and query planning cannot be effectively leveraged if the access has to be either a full consecutive record scan (as in the case of WARC file only) or recursive random access (for WARC and CDX).

If we are to implement all these potential optimizations, however, the effort will largely be duplicating many years of work already implemented by databases. To leverage the prior work, all that is required from us is to adopt the data formats already specified by databases. Optimized for query performance, database file formats have leveraged many sophisticated data structures, ranging from ordered flat files, ISAM, heap files, hash buckets, to B+ trees. Indeed, in the next section we will quantitatively evaluate analytics performance between data formatted in WARC and WARC/CDX with the same data set formatted in Parquet and Avro. The Parquet file format is designed to be used by columnar data stores while the Avro format is used by row stores. Both are open source formats, have many implementations, and already are widely used by big data systems, particularly Spark SQL.

Despite the performance penalty, WARC's data structure does carry an advantage over many others: it is very easy to append new records to existing ones. However, this advantage is only relevant for data collection. Files preserved for long term use are not expected to change over time, therefore easy appending is not an indispensable feature for an archival format.

## 4.2 Data Encoding: Text

Both WARC and CDX are encoded in pure text. This is a remnant of the upstream web messaging protocols. Still, it is worth questioning why a binary format, typically proven more efficient for both storage and processing than pure text, was not chosen for archiving. Afterall, web messaging protocols and archival formats serve drastically different purposes, so have different requirements. We can identify at least 3 major differences.

First, low interoperability and implementation overhead is the primary reason why web messaging protocols prefer text encoding. These protocols must be implemented by numerous software and systems. Each of them, acting as either a web client, server, or intermediary, is developed independently yet must be able to interact and coordinate seamlessly with each other. Messaging protocols specifying more sophisticated encodings, e.g., CORBA [26] or RPC [41], tend to attract far fewer implementers to surpass the critical mass to be as successful as the web. For web archives, however, interoperability is no longer essential since it only involves a small number of archives and their potential users. Like many other big data sources, a web archive can largely dictate the archival and/or dissemination formats as well as what is required for the potential users to access the data. Indeed, even though the support for WARC and/or CDX has been sparse for many years, they hold on as de facto web archival standard format, perhaps because the Internet Archive and many other national archives chose to preserve the web in these formats.

Second, text encoding is a major performance concern for batch web archival analytics but not for web messaging protocols. This is because the data granularity of web messaging is typically tiny – one HTTP request or response at a time. Further, the processing load on these small messages is distributed to many machines: each web browser is only responsible for analyzing and presenting the HTTP responses it receives, while each server is only responsible for parsing HTTP requests destined for it and providing HTTP responses originated from it. Nevertheless, DDoS attacks can still easily overwhelm web servers by wasting server resources on useless string parsing and database operations. In comparison, web archives disseminate data in fine granularity only under the close reading use case as illustrated by the Wayback Machine. Here, however, the Wayback Machine is responsible to disseminate responses for all archived URLs. Additionally, for batch web archival analytics workloads, we must be able to handle terabytes or even petabytes of data and all the required processing has to be shouldered by the users' analytical system. When batch processing textual files, we suspect that even tiny performance gains in either parsing, typecasting, or addressing, if multiplied by a large number, may result in significant speedups in end-to-end comparisons. This conjecture is corroborated by similar applications analyzing large JSON files or text logs. Two recent research studies estimated that they spend 80-90% of the execution time on parsing [32, 46]. In comparison, Freeman et al. [21] put this estimate to less than 40-70% when analyzing time-series data in binary encoding.

To understand why text encoding penalizes performance, let us take a string from a WARC file, "WARC-Date: 2006-09-19T17:20:24Z", as an example. It cannot be directly used by computers in a query, e.g., counting the number of records dated within a time range. We must first parse the whole WARC file to be able to separate one record from another and associate this string with a particular record. We also need to understand that this string is used to carry timestamp information, and tokenize it to isolate the substring "2006-09-19T17:20:24Z", and then typecast the substring into an integer signifying the number of seconds from an epoch time. We also need to typecast timestamps from all records before the comparison can be performed. In contrast, if binary encoding is allowed in a database file, as is the case for both Parquet and Avro, an integer is already stored in place for all timestamps in a specific database column. We will not need to parse, tokenize, and typecast, although in some cases we still need to load the full database into memory.

Third, textual encoding is chosen by web messaging protocols also for its easy human readability. This is useful for developers of diverse systems to debug errors. In web archiving use cases, however, it almost never occurs that a user needs to directly read an archived file. Even close reading is mediated by systems like the Wayback Machine.

In summary, if analytical performance is an important factor to choose a web archival file format, maintaining text encoding as part of the format specification is rather unnecessary.

## 4.3 Data Addressing: Lack Rules And Constraints For Efficient Addressing

Closely related to text encoding, WARC's addressing mechanism is also rudimentary. Here, addressing refers to the way a computer system pinpoints a particular piece of useful information from a large archival file, for example, the timestamp of a particular record. As explained in 4.2, when the file is text-encoded, we will have to parse the whole file, be it a WARC or CDX file. This does not pose a significant performance problem for web protocols because the data granularity is typically small and each message is independent from the others. When browsing the web, we never need to refer to the timestamp of a different HTTP transaction. Batch reusing web archives, however, involves many such references within a large file, and parsing the whole files for every query is expensive. Database files provide an efficient mechanism for addressing through the use of schema. With a schema, archival metadata will be designated various built-in or composite data types whose sizes are predetermined. For example, in Parquet all timestamps may be stored in INT64. The storage address of the $n$th record's timestamp can be easily computed without accessing any other field and record. Batch processing can therefore be more efficient by slashing unnecessary accesses.

## 4.4 Lost Opportunities for Combined Optimizations

When combining sophisticated data structure, encoding, and addressing, database files expose more opportunities for performance optimization. For example, Parquet may be queried so that large amounts of data accessing and processing are skipped, a mechanism known as Predicate Pushdown. These opportunities are not available for WARC and/or CDX.

## 4.5 Alternative Data Formats and Comparisons

In this section we briefly introduce two alternative data formats, Parquet and Avro, and compare their features with WARC/CDX. Figure 1 schematically illustrates their data structure differences.

Both Parquet and Avro file formats are natively supported by Spark. By default, Spark will use the Spark SQL dataframe object to load Parquet or Avro data instead of traditional RDD. With dataframes, users can view the loaded data as a table in a relational database and perform SQL queries against it. Through Spark SQL's catalyst optimizer, SQL queries also achieve better performance compared with similar programming language operations against RDD. This is because Spark SQL can correctly leverage the performance optimization opportunities offered by Parquet and Avro.

*4.5.1 Parquet.* Parquet is an open source columnar storage file format originally designed for the Hadoop filesystem by Cloudera and Twitter. It is optimized for fast querying, although writing Parquet files can be slow. This tradeoff is primarily attributed to its adoption of a sophisticated data structure. Following Google Dremel [38], Parquet rearranges nested data (e.g., XML and JSON) into multiple flat columns, each of which is made up of values of the same data type. For example, we can easily convert a CDX file into JSON without losing any information. We can then reformat the JSON into Parquet, which will store all timestamps into a single column of INT64 for easier and faster query.

In addition to the WORM (write once read many) support described above, Parquet's columnar storage also provides significant performance advantages for batch processing workloads:

- High compression. The unique data organization within Parquet can make the compression more efficient, which can reduce the data size and improve serialization performance inside the Spark framework. As shown in section 5, we are able to achieve 0.9 compression ratio when compared with WARC, both with GZIP compression encoding. We also achieve 1.5x speed up on full data scan compared with WARC.
- Minimize data input with predicate pushdown. Data in Parquet is first divided by a number of row groups, and each row group is split into chunks of columns. Parquet then embeds extra information in the file, e.g., min/max stats and dictionary for column values in each row group. For queries like "count number of records with timestamp within a certain range", if the Parquet file is sorted by timestamp, many row groups will have timestamp min/max stat outside of the query range. Querying systems can therefore skip loading these row groups altogether, a mechanism known as predicate pushdown.
- Minimize data input with column projection pushdown. Columns in Parquet are stored separately and only relevant columns will be accessed during the query. For example, the aforementioned time range query does not need to know anything about the payload. Therefore, like in WARC/CDX, payload data in Parquet do not need to be accessed. Furthermore, since the query does not concern any other columns except for "timestamp", all other metadata columns will not be accessed. This is also known as projection pushdown. Figure 2 schematically illustrates the effects of both predicate pushdown and projection pushdown. In our evaluation, when using Spark SQL to query Parquet, the combination of these two pushdowns can achieve speedup up to two orders of magnitude. Querying a Parquet file can also be around 2 times faster than querying CDX with the same data.

*4.5.2 Avro.* Avro is another open source row-based file format, developed by the Hadoop project. Avro is optimized for data serialization and schema evolution. Similar to WARC/CDX, Avro is highly splittable and concatenatable: each record is preserved in one block with associate sync marker (similar to offset), and the sync marker can be used for random access. Different from WARC, however, Avro adopts JSON data types and uses binary encoding. Data addressing in Avro depends heavily on the schema. Avro performs the best for batch web archival processing that needs to access lots of payload data. In full data scan, Avro can achieve 2X speedup compared with WARC and 1.6X compared with Parquet.

## 5 EVALUATION

To benchmark the qualitative analysis in section 4, we conduct controlled experiments. Because our main purpose is to evaluate the performance impact of the data format, we reformat the original WARC data set in different ways but they all contain the same information, albeit arranged differently. We then apply representative analytical workloads to them to trace the root cause of the performance differences.

### 5.1 Hardware/Software Setup

Experiments are performed on a shared-nothing cluster consisting of 1 master node and 6 compute nodes, inter-connected with an HP 2910al-48G gigabit Ethernet switch. Each node is a Supermicro 6027TR-DTRF server with 2 x 8-core Intel® Xeon® E5-2660 CPU, 64 GiB Memory, and 4 x 4 TB SATA hard disk. The cluster runs Cloudera Distribution Hadoop (CDH) 6.3.0, with Hadoop 3.0.0 and Spark 2.4.0. Data files are copied to HDFS. All experiments are executed in Spark Read Evaluate Print Loop (REPL) environment through Apache Zeppelin notebook. We use the following Spark configurations for all Spark jobs: yarn-client mode, 5 executor cores, and 30 Gb executor memory.

### 5.2 Data

We chose to use Common Crawl's web archiving data crawled from May 20 to 23, 2018. The data set consists of 1219 gzip compressed WARC files totaling 0.98 TB, and contains 53324440 records. The WARC files are organized by crawling time, each containing records crawled from a mutually exclusive time span.

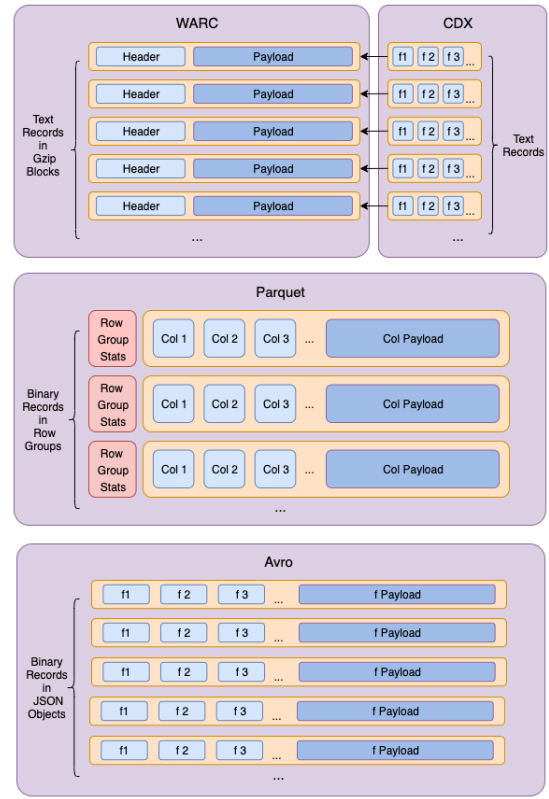

Figure 1: Comparing WARC-CDX, Parquet, and Avro formats

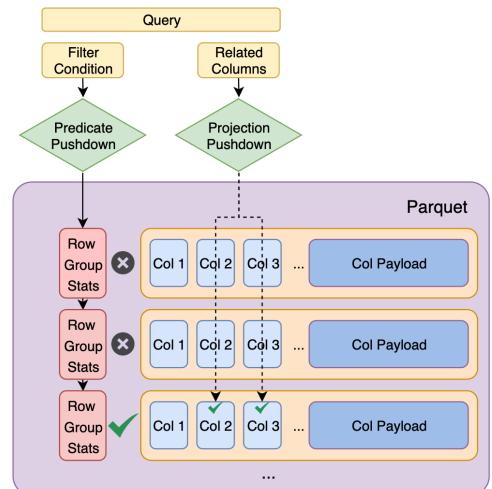

Figure 2: Use predicate pushdown and projection pushdown to skip reading unnecessary row groups or columns in Parquet data

We then reformat the WARC files to yield the following five datasets for comparison: 1) the original WARC files; 2) case 1 plus CDX index files built against all the original WARC files; 3) Parquet files containing the same information as case 1, with most columns in String type; 4) the same as case 3 but the Timestamp column is in INT64 Timestamp type; 5) Avro, with the same column data types as in case 4 but the Timestamp column is in INT96.

|  | WARC | WARC-CDX | Avro | Parquet |
|---|---|---|---|---|
| Average Runtime (sec) | 2891 | 33 | 1245 | 15 |

Table 1: Task 1: Count the number of records

|  | WARC | WARC-CDX | Avro | Parquet | Parquet (No-PP) |
|---|---|---|---|---|---|
| Max | 3366 | 36.70 | 1493 | 17.07 | 40.14 |
| Med | 3357 | 35.28 | 1450 | 15.86 | 35.43 |
| Min | 3354 | 34.51 | 1427 | 15.49 | 31.59 |

Table 2: Task 2: Retrieve metadata from records filtered by a given time range (Runtime in sec)

We use a modified ArchiveUnleashedToolkit for reformating WARC to Parquet and Avro. Settings for Parquet/Avro file writing are included in the example code repository on Github [2]. The reformatted Parquet and Avro data are gzip compressed, the same as WARC files.

### 5.3 Workloads

We perform the following analytical tasks, with Task 1 to 5 against the 5 datasets specified in section 5.2, and Task 6 against 5 filtered datasets created from the previous 5 in the same formats. The filtered dataset is about 5GB when formatted in WARC.

- Task 1: Count the number of records
- Task 2: Retrieve metadata from records filtered by a given time range
- Task 3: Retrieve metadata from records filtered by a list of URLs
- Task 4: Retrieve full records filtered by a given time range
- Task 5: Retrieve full records filtered by a list of URLs
- Task 6: Given a dataset, extract pure text from all payload and run topic modeling over the extracted text records

Queries and example codes used in the experiments can found on Github[2].

Tasks 1 to 3 are Type 1 workloads that only need metadata to complete. Tasks 4 and 5 are Type 2 workloads that will load all filtered records into the memory. Task 6 is the Type 3 workload that will load all records in the dataset into the memory.

For tasks 2 to 5, we also use different time ranges or URL lists to yield different levels of selectivity from the dataset. A selectivity of 0.001 means the query has yielded 0.1% of the total number of records. When we vary the time ranges on Task 2, however, the results do not change with selectivity. We therefore do not plot the results against selectivity.

For Type 1 work loads (tasks 1 to 3), the query results only include metadata. For Type 2 work loads (tasks 4 and 5), the query results include the full record including payload content.

Type 3 workload (task 6) involves retrieving the payload content from the collection, extracting pure textual content from the payload, and applying the topic model algorithm to derived text content. In our experiments, we use a subset of about 5GB collection to represent a derived web collection. We use the Latent Dirichlet Allocation (LDA) algorithm from SparkMLib [39] for topic modeling.

### 5.4 Results and Analysis

As shown in Table 1, Table 2 and Figure 3, for all Type 1 workloads, Parquet performs the best. This may be attributed to the columnar data structure that allows Spark SQL to skip loading irrelevant columns, which significantly reduces the disk I/O. In task 2, the difference between Parquet with no predicate pushdown (No-PP), where timestamp is encoded in String, and Parquet, where Timestamp is encoded in INT64, illustrates the performance gain from the binary encoding as well as Predicate Pushdown. Parquet and WARC-CDX performance are significantly better than Avro and WARC. This is because when queries only concern metadata, only the CDX file and the relevant column(s) in Parquet will be accessed. In contrast, all data contained in WARC and Avro will need to be loaded into memory.

[2]https://github.com/xw0078/WebArchiveWithParquetAvro

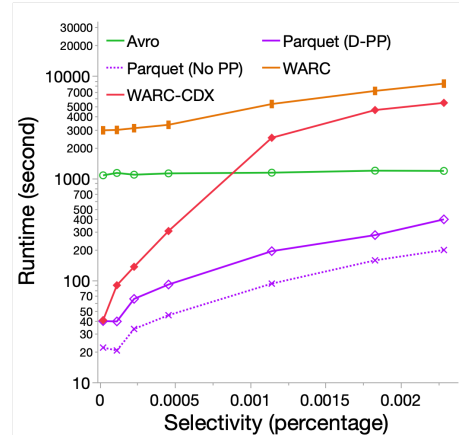

Figure 3: Task 3: Retrieve metadata from records filtered by a list of URLs

As shown in Figures 4 and 6, Parquet still demonstrates strong performance in Type 2 workloads. At low selectivity, the effects of Predicate and Projection Pushdown are more evident, allowing it to achieve about 2 orders of magnitude of performance gain against WARC. WARC-CDX is only faster when the selectivity is very low. Since Type 2 workloads need to load payloads from many records into the memory yet WARC-CDX uses recursive random disk access, its performance quickly degrades and will continue to worsen as selectivity goes higher. Figure 4 shows that when the selectivity is above 25%, WARC-CDX's performance is worse than reading all WARC files in full. The CDX index becomes counterproductive. In contrast, the Parquet's performance degradation converges to Parquet (No-PP).

For Type 3 workload (task 6), the results demonstrate the disadvantage of recursive random access posed by WARC-CDX. The LDA algorithm needs to access all payload data. Under WARC-CDX, applications will need to query the CDX file for addresses of each record then repeatedly copy these records from disk to memory one at a time. Under WARC, however, the whole file may be copied in at one time. The disk I/O efficiency afforded by database file formats is now at full display, making the Parquet and Avro about one order of magnitude faster than WARC-CDX. While WARC-CDX is relatively fast for small metadata queries, it is not quite competent for batch processing. Figure 4 also compares the data size of these formats (all are gzip compressed except for CDX). Spatial efficiency wise, AVRO is slightly worse than the others, but the difference may warrant trading it for WARC's much worse batch processing performance.

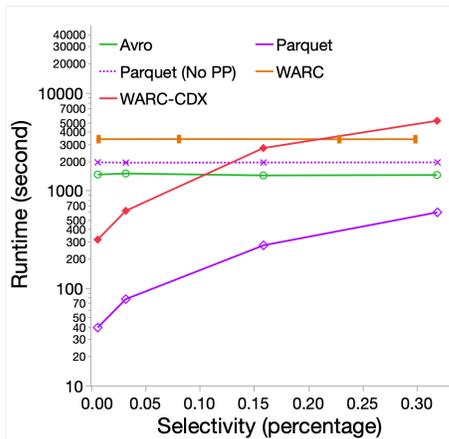

Figure 4: Task 4: Retrieve full records filtered by a given time range

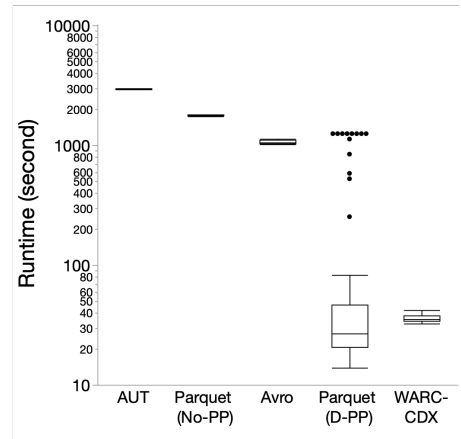

Figure 6: Task 5 with a single URL

|  | WARC | WARC-CDX | Avro | Parquet |
|---|---|---|---|---|
| Size (TB) | 0.985 | 0.998 | 1.321 | 0.914 |

Table 4: Data Size

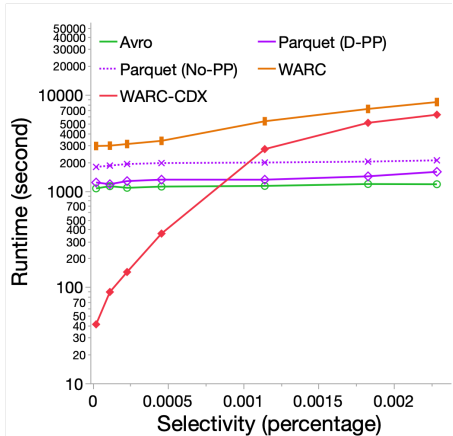

Figure 5: Task 5: Retrieve full records filtered by a list of URLs

|  | WARC | WARC-CDX | Avro | Parquet |
|---|---|---|---|---|
| Average Runtime (sec) | 5450 | 2115 | 466 | 484 |

Table 3: Task 6: Topic modeling

In summary, our evaluation demonstrates the performance advantages to format web archival data in optimized database formats such as Parquet or Avro.

### 5.5 Transactional Work Load

While our experiments clearly demonstrate the WARC-CDX's performance degradation for batch data processing, we are also interested to gauge its advantage on small, transactional queries. Figure 6 shows the result of Task 5 filtered against a single URL. This is similar to the workload of the Wayback Machine. On average, Parquet with Predicate Pushdown on domain name (D-PP) is actually faster than WARC-CDX. However, in rare cases, Predicate Pushdown does not work very well. In these cases, the different captures of a single URL may have been evenly distributed in many row groups, forcing the application to read all these row groups, thus the performance degradation.

Another point to note is that the Parquet file we created sort the row groups by Timestamp. Predicate Pushdown therefore works better on Timestamp filtering than URL filtering, as shown from Figures 4 and 6. If URL filtering performance is more important, we can always re-sort the row groups differently, e.g., based on domain name, although there is a limit on how much we can optimize for all types of queries. Nevertheless, as Figure 11 clearly illustrates, even sub-optimal performance from Parquet can be comparable to or outperform WARC-CDX on transactional workloads.

## 6 CONCLUSIONS AND DISCUSSIONS

As we point out in section 4, WARC's resemblances to web messaging allow crawlers to defer many data loading, parsing, conversion, and validation workloads during the data collection stage. However, if we choose to also use WARC as the archival format and eliminate such processing altogether, the end users of the archive will have to shoulder extra processing burden at the reuse stage. Our evaluations provide evidences that WARC carries significant performance penalties for batch data processing workloads, up to two orders of magnitude slower than that of more efficient formats. We therefore call for the web archiving community to consider adopting alternative archival formats.

On the other hand, our recommendation does not necessarily imply the end of life for WARC. In terms defined by the OAIS reference model [11], the WARC format remains valuable when collecting data to build Submission Information Packages (SIPs), especially when resources are limited during the crawling. But more efficient formats should be considered for Archival Information Packages (AIPs). The format conversion from SIPs to AIPs may be carried out offline when computing and storage resources are no longer a bottleneck. Indeed this is already happening. A number of WARC derivative formats have been proposed to store post-processing results. For example, CDX/CDXJ [14] are plain text formats used to store HTTP headers [12]; WAT is a JSON based format used to store WARC metadata [22]; WET is again a plain text format used to store extracted textual elements [40]. More recently, Common Crawl even directly used Parquet to store metadata extracted from WARC [44]. The weakness of this approach, however, is its dependency on an inefficient storage format. Because ArchiveSpark [27] is the only published example we are aware of that leverages a derivative format in batch web archive analysis, we use it as a surrogate to evaluate the performance impact of WARC derivatives. The results show the performance is only comparable to more efficient formats for transactional workloads. For Type 2 and 3 batch workloads, because WARC derivative formats can only store record-level offsets instead of directly managing the data storage, the performance is throttled by the addressing mechanism that has been demonstrated by our evaluations to be inefficient. We anticipate this trend to hold true even if the metadata are stored in Parquet as in [44]. Even for Type 1 batch workloads where the absolute difference between querying a pure text metadata file and querying a database file is not as pronounced, the 2X to 3X relative difference can still be significant with the growth of web archives. It is therefore not sustainable to continue storing metadata in pure text formats, as is the case for CDX, WAT, and WET.

As with many other discussions around data formats, arguments can be made to emphasize the simplicity, flexibility, and durability of a text-based format such as WARC and its derivatives against a binary format. However, many such arguments need to be put in the context of big data and its usage. Not only does human readability factor far less significantly than machine efficiency with the data growth in size, many of its perceived advantages have also been rendered irrelevant. For example, almost all archived WARC files we encountered have already been compressed and converted into binaries in the archival storage. The notion that a human-readable text is more durable because it can sustain benign neglect for an extended period of time, in our view, lacks quantitative support. Qualitatively, contrary arguments can also be made that free-form natural languages appear to be harder to decode and process than machine languages, whose semantics and syntax are mechanical thus simpler. Specifically, a database format can be decoded with the associated schema and the schema can be extended to include unexpected elements in the future.

Moreover, neither Parquet nor Avro depends on the existence of HDFS and/or Spark to function as an archival format. They simply offer more optimized layout and hooks to leverage data sharding, replication, and parallel processing provided by these systems. Even without taking full advantages of these optimizations, the performance is not necessarily worse off than analyzing flat textual files, which offer little opportunity to speed up queries. The support for open database formats can also be more available and sustained than that of a niche text format. Moreover, the loss of abandoning many years of work around WARC can in turn be compensated by leveraging a much larger body of established work in database, distributed systems, and parallel processing. The performance gain will open the web archives for more productive and effective reuse.

Our recommendation is also inline with the academic and industrial trends in big data processing. While flat text log format continues to be used to collect data and JSON used by APIs to disseminate data, they have long been considered a major bottleneck for data warehousing and analytics. Much work has been done and new techniques developed to speed up the parsing and loading of log files and JSON as well as to convert them into more efficient formats for batch processing. A recent example from Uber [51] describes how the company incrementally converts its JSON source data into Parquet and Avro in a 100+ Petabyte data lake with minute level query latency. In order to achieve similar performance criteria at a web archive, we think it is necessary to move away from using WARC as the archival format.

## ACKNOWLEDGMENTS

This work is partially supported by the Institute of Museum and Library Services under Grant No.: LG-71-16-0037-16 (https://www.imls.gov/grants/awarded/lg-71-16-0037-16) and National Science Foundation under grant IIS-1619028 (https://www.nsf.gov/awardsearch/showAward?AWD_ID=1619028) and IIS-1619371 (https://www.nsf.gov/awardsearch/showAward?AWD_ID=1619371).

We also thank anonymous reviewers who have provided helpful comments on the submitted draft of the manuscript.

## REFERENCES

[1] Ioannis Alagiannis, Renata Borovica-Gajic, Miguel Branco, et al. 2015. NoDB: Efficient Query Execution on Raw Data Files. *Commun. ACM* 58, 12 (Nov. 2015), 112–121. https://doi.org/10.1145/2830508

[2] Sawood Alam, Mat Kelly, and Michael L. Nelson. 2016. InterPlanetary Wayback: The Permanent Web Archive. In *Proceedings of the 16th ACM/IEEE-CS Joint Conference on Digital Libraries (JCDL '16)*. ACM, Newark, New Jersey, USA, 273–274. https://doi.org/10.1145/2910896.2925467

[3] Sawood Alam, Michael L. Nelson, Herbert Van de Sompel, et al. 2016. Web Archive Profiling through CDX Summarization. *International Journal on Digital Libraries* 17, 3 (Sept. 2016), 223–238. https://doi.org/10.1007/s00799-016-0184-4

[4] Sawood Alam, Michele Weigle, Michael Nelson, et al. 2019. MementoMap Framework for Flexible and Adaptive Web Archive Profiling. In *2019 ACM/IEEE Joint Conference on Digital Libraries (JCDL)*. IEEE, 172–181. https://doi.org/10.1109/JCDL.2019.00033

[5] Fernando Amat, Burkhard Höckendorf, Yinan Wan, et al. 2015. Efficient Processing and Analysis of Large-Scale Light-Sheet Microscopy Data. *Nature Protocols* 10, 11 (Nov. 2015), 1679–1696. https://doi.org/10.1038/nprot.2015.111

[6] Apache. 2020. Apache/Parquet-Format. The Apache Software Foundation. Retrieved Jan. 20, 2020 from https://github.com/apache/parquet-format

[7] Apache. 2020. Avro. The Apache Software Foundation. Retrieved Jan. 20, 2020 from https://github.com/apache/avro

[8] Internet Archive. 2020. Internet Archive. Retrieved Jan. 20, 2020 from https://archive.org/

[9] Internet Archive. 2020. Wayback Machine. Retrieved Jan. 20, 2020 from https://archive.org/web/

[10] Niels Brügger and Ian Milligan (Eds.). 2019. *The SAGE Handbook of Web History* (first ed.). SAGE Publications Ltd.

[11] CCSDS. 2012. *Space Data and Information Transfer Systems — Open Archival Information System (OAIS) — Reference Model*. Standard ISO 14721:2012. International Organization for Standardization.

[12] International Internet Preservation Consortium. 2020. The CDX File Format. Retrieved Jan. 20, 2020 from https://iipc.github.io/warc-specifications/specifications/cdx-format/cdx-2006/

[13] International Internet Preservation Consortium. 2020. Openwayback. Retrieved Jan. 20, 2020 from https://github.com/iipc/openwayback

[14] International Internet Preservation Consortium. 2020. OpenWayback CDXJ File Format. Retrieved Jan. 20, 2020 from https://iipc.github.io/warc-specifications/specifications/cdx-format/openwayback-cdxj/

[15] Miguel Costa, Daniel Gomes, and Mário J. Silva. 2017. The Evolution of Web Archiving. *International Journal on Digital Libraries* 18, 3 (Sept. 2017), 191–205. https://doi.org/10.1007/s00799-016-0171-9

[16] Common Crawl. 2020. Common Crawl. Retrieved Jan. 20, 2020 from https://commoncrawl.org/

[17] H. Van de Sompel, M. Nelson, and R. Sanderson. 2013. *HTTP Framework for Time-Based Access to Resource States – Memento*. RFC 7089. Internet Engineering Task Force. https://doi.org/10.17487/RFC7089
[18] Zeon Trevor Fernando, Ivana Marenzi, and Wolfgang Nejdl. 2018. ArchiveWeb: Collaboratively Extending and Exploring Web Archive Collections—How Would You like to Work with Your Collections? *International Journal on Digital Libraries* 19, 1 (March 2018), 39–55. https://doi.org/10.1007/s00799-016-0206-2
[19] Roy T. Fielding, James Gettys, Jeffrey C. Mogul, et al. 1999. *Hypertext Transfer Protocol – HTTP/1.1*. RFC 2616. Internet Engineering Task Force. https://doi.org/10.17487/RFC2616
[20] Ned Freed and Nathaniel S. Borenstein. 1996. *Multipurpose Internet Mail Extensions (MIME) Part Two: Media Types*. RFC 2046. Internet Engineering Task Force. https://doi.org/10.17487/RFC2046
[21] Jeremy Freeman, Nikita Vladimirov, Takashi Kawashima, et al. 2014. Mapping Brain Activity at Scale with Cluster Computing. *Nature Methods* 11, 9 (Sept. 2014), 941–950. https://doi.org/10.1038/nmeth.3041
[22] Vinay Goel. 2011. Web Archive Metadata File Specification. Retrieved Jan. 20, 2020 from https://webarchive.jira.com/wiki/spaces/Iresearch/pages/13467719/Web+Archive+Metadata+File+Specification
[23] Daniel Gomes, João Miranda, and Miguel Costa. 2011. A Survey on Web Archiving Initiatives. In *Research and Advanced Technology for Digital Libraries*, Stefan Gradmann, Francesca Borri, Carlo Meghini, and Heiko Schuldt (Eds.). Springer Berlin Heidelberg, 408–420.
[24] Shawn Graham, Ian Milligan, and Scott Weingart. 2015. *Exploring Big Historical Data: The Historian's Macroscope* (reprint ed.). Imperial College Press, London.
[25] P. Greenfield, M. Droettboom, and E. Bray. 2015. ASDF: A New Data Format for Astronomy. *Astronomy and Computing* 12 (Sept. 2015), 240–251. https://doi.org/10.1016/j.ascom.2015.06.004
[26] Object Management Group. 2012. *Common Object Request Broker Architecture*. Standard 3.3. https://www.omg.org/spec/CORBA/3.3/
[27] Helge Holzmann, Vinay Goel, and Avishek Anand. 2016. ArchiveSpark: Efficient Web Archive Access, Extraction and Derivation. In *2016 ACM/IEEE Joint Conference on Digital Libraries (JCDL)*. ACM, New York, NY, USA, 83–92. https://doi.org/10.1145/2910896.2910902
[28] ISO. 2009. *Information and Documentation - WARC File Format*. Standard ISO 28500:2009. International Organization for Standardization.
[29] Ian Jacobs and Norman Walsh. 2004. Architecture of the World Wide Web, Volume One. W3C Recommendation 15 December 2004. *World Wide Web Consortium* (2004). http://www.w3.org/TR/webarch/
[30] Manos Karpathiotakis, Miguel Branco, Ioannis Alagiannis, et al. 2014. Adaptive Query Processing on RAW Data. *Proc. VLDB Endow.* 7, 12 (Aug. 2014), 1119–1130. https://doi.org/10.14778/2732977.2732986
[31] Martin Klein, Herbert Van de Sompel, Robert Sanderson, et al. 2014. Scholarly Context Not Found: One in Five Articles Suffers from Reference Rot. *PLOS ONE* 9, 12 (Dec. 2014), 115253. https://doi.org/10.1371/journal.pone.0115253
[32] Yinan Li, Nikos R. Katsipoulakis, Badrish Chandramouli, et al. 2017. Mison: A Fast JSON Parser for Data Analytics. *Proc. VLDB Endow.* 10, 10 (June 2017), 1118–1129. https://doi.org/10.14778/3115404.3115416
[33] Jimmy Lin, Milad Gholami, and Jinfeng Rao. 2014. Infrastructure for Supporting Exploration and Discovery in Web Archives. In *Proceedings of the 23rd International Conference on World Wide Web*. ACM, New York, NY, USA, 851–856. https://doi.org/10.1145/2567948.2579045
[34] Jimmy Lin, Ian Milligan, Jeremy Wiebe, et al. 2017. Warcbase: Scalable Analytics Infrastructure for Exploring Web Archives. *J. Comput. Cult. Herit.* 10, 4 (July 2017), 22:1–22:30. https://doi.org/10.1145/3097570
[35] Jialin Liu, Evan Racah, Quincey Koziol, et al. 2016. H5Spark: Bridging the I/O Gap between Spark and Scientific Data Formats on HPC Systems. In *CUG2016 Proceedings*. Cray User Group, London, England, UK.
[36] Madhavi Mallapragada. 2019. Cultural Historiography of the 'Homepage'. In *The SAGE Handbook of Web History*. SAGE Publications Ltd, 387–399. https://doi.org/10.4135/9781526470546
[37] Matt Massie, Frank Nothaft, Christopher Hartl, et al. 2013. *Adam: Genomics Formats and Processing Patterns for Cloud Scale Computing*. University of California, Berkeley Technical Report UCB/EECS-2013. UCB/EECS-2013-207, EECS Department, University of California, Berkeley.
[38] Sergey Melnik, Andrey Gubarev, Jing Jing Long, et al. 2010. Dremel: Interactive Analysis of Web-Scale Datasets. *Proc. VLDB Endow.* 3, 1-2 (Sept. 2010), 330–339. https://doi.org/10.14778/1920841.1920886
[39] Xiangrui Meng, Joseph Bradley, Burak Yavuz, et al. 2015. MLlib: Machine Learning in Apache Spark. (May 2015). arXiv:1505.06807
[40] Stephen Merity. 2014. Navigating the WARC File Format. Library Catalog: commoncrawl.org.
[41] Sun Microsystems. 1988. *RPC: Remote Procedure Call Protocol specification*. RFC 1050. Internet Engineering Task Force. https://doi.org/10.17487/RFC1050
[42] Ian Milligan. 2019. *History in the Age of Abundance?: How the Web Is Transforming Historical Research*. McGill-Queen's University Press, Montreal.
[43] Jessica D. Mink. 2015. Astronomical Data Formats: What We Have and How We Got Here. *Astronomy and Computing* 12 (Sept. 2015), 128–132. https://doi.org/10.1016/j.ascom.2015.07.001
[44] Sebastian Nagel. 2018. Index to WARC Files and URLs in Columnar Format. Retrieved Jan. 20, 2020 from https://commoncrawl.org/2018/03/index-to-warc-files-and-urls-in-columnar-format/
[45] Frank Austin Nothaft, Matt Massie, Timothy Danford, et al. 2015. Rethinking Data-Intensive Science Using Scalable Analytics Systems. In *Proceedings of the 2015 ACM SIGMOD International Conference on Management of Data*. ACM, Melbourne, Victoria, Australia, 631–646. https://doi.org/10.1145/2723372.2742787
[46] Shoumik Palkar, Firas Abuzaid, Peter Bailis, et al. 2018. Filter Before You Parse: Faster Analytics on Raw Data with Sparser. *Proc. VLDB Endow.* 11, 11 (July 2018), 1576–1589. https://doi.org/10.14778/3236187.3236207
[47] Andrew Pavlo, Erik Paulson, Alexander Rasin, et al. 2009. A Comparison of Approaches to Large-Scale Data Analysis. In *Proceedings of the 35th SIGMOD International Conference on Management of Data - SIGMOD '09*. ACM Press, Providence, Rhode Island, USA, 165. https://doi.org/10.1145/1559845.1559865
[48] Julien Peloton, Christian Arnault, and Stéphane Plaszczynski. 2018. FITS Data Source for Apache Spark. *Computing and Software for Big Science* 2, 1 (Oct. 2018), 7. https://doi.org/10.1007/s41781-018-0014-z
[49] P. Resnick. 2001. *Internet Message Format*. RFC 2822. Internet Engineering Task Force. https://doi.org/10.17487/RFC2822
[50] Hany M. SalahEldeen and Michael L. Nelson. 2012. Losing My Revolution: How Many Resources Shared on Social Media Have Been Lost?. In *Theory and Practice of Digital Libraries (Lecture Notes in Computer Science)*, Panayiotis Zaphiris, George Buchanan, Edie Rasmussen, and Fernando Loizides (Eds.). Springer, Berlin, Heidelberg, 125–137. https://doi.org/10.1007/978-3-642-33290-6_14
[51] Reza Shiftehfar. 2018. Uber's Big Data Platform: 100+ Petabytes with Minute Latency. Retrieved Jan. 20, 2020 from https://eng.uber.com/uber-big-data-platform/
[52] Archives Unleashed Team. 2020. Archivesunleashed/Aut. Archives Unleashed. Retrieved Jan. 20, 2020 from https://github.com/archivesunleashed/aut
[53] B. Thomas, T. Jenness, F. Economou, et al. 2015. Learning from FITS: Limitations in Use in Modern Astronomical Research. *Astronomy and Computing* 12 (2015), 133–145. https://doi.org/10.1016/j.ascom.2015.01.009
[54] Herbert Van de Sompel, Michael L. Nelson, Robert Sanderson, et al. 2009. Memento: Time Travel for the Web. (2009). arXiv:0911.1112
[55] Matthew S. Weber. 2018. Methods and Approaches to Using Web Archives in Computational Communication Research. *Communication Methods and Measures* 12, 2-3 (April 2018), 200–215. https://doi.org/10.1080/19312458.2018.1447657
[56] Webrecorder. 2020. Pywb. Webrecorder. Retrieved Jan. 20, 2020 from https://github.com/webrecorder/pywb
[57] Zhiwu Xie and Edward A. Fox. 2017. Advancing Library Cyberinfrastructure for Big Data Sharing and Reuse. *Information Services & Use* 37, 3 (Jan. 2017), 319–323. https://doi.org/10.3233/ISU-170853
[58] Tian Yang, Kenjiro Taura, and Liu Chao. 2017. SDAC: Porting Scientific Data to Spark RDDs. In *Network and Parallel Computing (Lecture Notes in Computer Science)*, Xuanhua Shi, Hong An, Chao Wang, Mahmut Kandemir, and Hai Jin (Eds.). Springer International Publishing, 127–130.
[59] Hao Zhang, Gang Chen, Beng Chin Ooi, et al. 2015. In-Memory Big Data Management and Processing: A Survey. *IEEE Transactions on Knowledge and Data Engineering* 27, 7 (July 2015), 1920–1948. https://doi.org/10.1109/TKDE.2015.2427795